\documentclass[aps,prb,floatfix,twocolumn,showpacs]{revtex4}%
\usepackage{amsmath}
\usepackage{graphicx}
\usepackage{amsfonts}
\usepackage{amssymb}%
\providecommand{\U}[1]{\protect\rule{.1in}{.1in}}
\providecommand{\U}[1]{\protect\rule{.1in}{.1in}}
\providecommand{\U}[1]{\protect\rule{.1in}{.1in}}
\providecommand{\U}[1]{\protect\rule{.1in}{.1in}}
\providecommand{\U}[1]{\protect\rule{.1in}{.1in}}
\providecommand{\U}[1]{\protect\rule{.1in}{.1in}}

\begin{document}
\title{Squeezing light with Majorana fermions}
\author{Audrey Cottet$^{1}$ , Takis Kontos$^{1}$ and Benoit Dou\c{c}ot$^{2}$}
\affiliation{$^{1}$Laboratoire Pierre Aigrain, Ecole Normale Sup\'{e}rieure, CNRS UMR 8551,
Laboratoire associ\'{e} aux universit\'{e}s Pierre et Marie Curie et Denis
Diderot, 24, rue Lhomond, 75231 Paris Cedex 05, France}
\affiliation{$^{2}$Laboratoire de Physique Th\'{e}orique et des Hautes Energies, CNRS UMR
7589, Universit\'{e}s Paris 6 et 7, 4 Place Jussieu, 75252 Paris Cedex 05}
\date{\today}


\begin{abstract}
Coupling a semiconducting nanowire to a microwave cavity provides a powerfull
means to assess the presence or absence of isolated Majorana fermions in the
nanowire. These exotic bound states can cause a significant cavity frequency
shift but also a strong cavity nonlinearity leading for instance to light
squeezing. The dependence of these effects on the nanowire gate voltages gives
direct signatures of the unique properties of Majorana fermions, such as their
self-adjoint character and their exponential confinement.

\end{abstract}
\maketitle

\section{Introduction}

\begin{figure}[t]
\includegraphics[width=1.\linewidth]{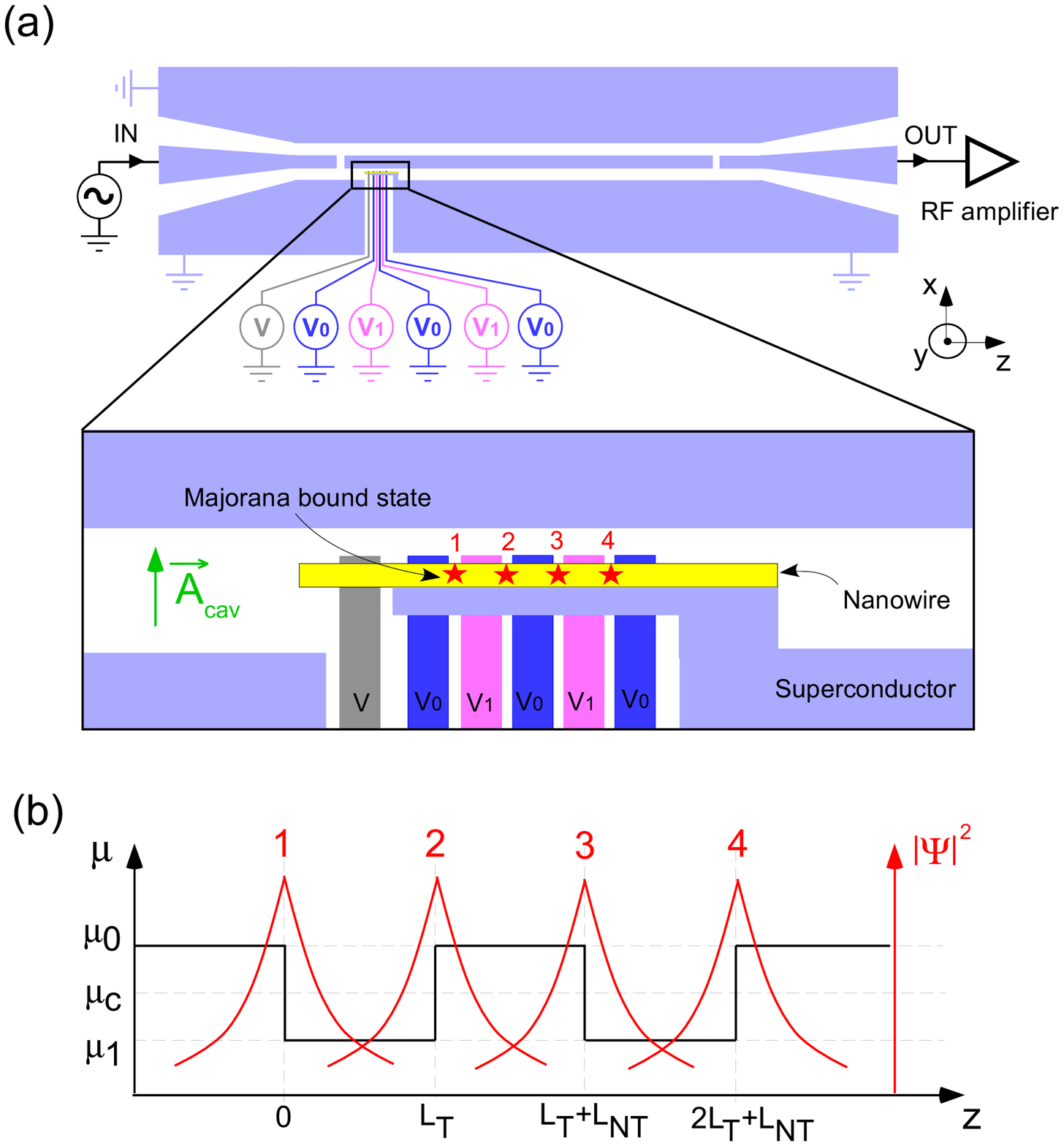}\caption{a.\textbf{ }Scheme our
our setup. The microwave cavity is made from the various superconducting
contacts in purple. The nanowire (yellow) is placed between the center and
ground conductors of the cavity. It is tunnel contacted to a grounded
superconducting contact (purple) and capacitively contacted to three gate
electrodes with voltage $V_{0}$ (blue) and two gate electrodes with voltage
$V_{1}$ (pink) used to impose chemical potentials $\mu_{0}$ and $\mu_{1}$ in
different sections of the nanowire. A normal metal contact (grey) with bias
voltage $V$ is tunnel contacted to the nanowire to perform a conductance
spectroscopy. b. Chemical potential (left axis) and schematic quasiparticle
density probability (right axis) in the nanowire, versus coordinate $z$. Four
MBSs appear at the boundaries between the topological ($\mu=\mu_{1}$) and
non-topological ($\mu=\mu_{0}$) sections of the nanowire. In a finite length
system, the MBSs\ wavefunctions overlap. }%
\end{figure}

The observation of isolated Majorana fermions in hybrid nanostructures is one
of the major challenges in quantum electronics. These elusive quasiparticles
borrowed from high energy physics have the remarkable property of being their
own antiparticle\cite{Majorana}. They are expected to appear as zero energy
localized modes in various types of heterostructures\cite{reviews}. One
promising strategy is to use semiconducting nanowires with a strong spin-orbit
coupling, such as InAs and InSb nanowires, placed in proximity with a
superconductor and biased with a magnetic field \cite{Lutchyn,Oreg}. Most of
the recent experiments proposed and carried out have focused on electrical
transport which appears as the most natural probe in electronic
devices\cite{Lutchyn,Oreg,Bolechetc,Flensberg}. While signatures consistent
with the existence of Majorana fermions have been observed
recently\cite{Mouriketc}, it is now widely accepted that alternative
interpretations can explain most of the experimental findings observed so
far\cite{Lee,Liu,Pikulin,Rainis,Zareyan,CottetSF,Sau}. One has therefore to do
more than the early transport experiments to demonstrate unambiguously the
existence of Majorana particles in condensed matter. Here, we propose to use
the tools of cavity quantum electrodynamics to perform this task. Photonic
cavities, or generally harmonic oscillators, are extremely sensitive detectors
which can be used to probe fragile light-matter hybrid coherent
states\cite{Wallraff:04}, non-classical light or even possibly gravitational
waves\cite{Clerk}. We show here that a photonic cavity can also be used to
detect Majorana fermions and test their unique properties.

Recent technological progress has enabled the fabrication of nanocircuits
based for instance on InAs nanowires inside coplanar microwave
cavities\cite{Delbecq,DQD,Petersson}. On the theory side, it has been
suggested to couple nanowires to cavities to produce Majorana polaritons
\cite{Trif}, or build qubit architectures\cite{Trif2}. Here, we adopt a
different perspective which is the direct characterization of Majorana bound
states (MBSs) through a photonic cavity. We consider a nanowire with four well
defined Majorana bound states (MBSs), away from the nanowire topological
transition. We find that these MBSs can be strongly coupled to the cavity when
their spatial extension is large enough. When the four MBSs are coupled to the
cavity, this leads to a transverse coupling scheme which induces a cavity
frequency shift but also strong nonlinearities in the cavity behavior, such as
light squeezing\cite{Ong,Kirchmair}. Using electrostatic gates, it is possible
to reach a regime where only two MBS remain coupled to the cavity. In this
case, the cavity frequency shift and nonlinearity disappear. This represents a
direct signature of the particle/antiparticle duality of MBSs. Indeed, the
self-adjoint character of MBSs forces a longitudinal coupling to the cavity
when only two MBSs are coupled to the cavity. The evolution of the cavity
frequency shift and nonlinearity with the nanowire gate voltages furthermore
enables an almost direct observation of the exponential localization of MBSs.

This article is organized as follows. In section II, we present the low-energy
Hamiltonian model of the 4 Majorana nanowire considered in this article. In
section III, we discuss the tunnel spectroscopy of this nanowire, through a
normal metal contact placed close to one of the Majorana bound states. In
section IV, we discuss the coupling between the nanowire and a microwave
cavity. In section V, we discuss the behavior of the microwave cavity in the
dispersive regime where the Majorana system and the cavity are not resonant.
In section VI we discuss various simplifications used in our approach. Section
VII concludes. For clarity, we have postponed various technical details and
calculations to appendices. Appendix A presents a one-dimensional microscopic
description of the nanowire, used to obtain the parameters occurring in the
low energy Hamiltonian of section II and the coupling between the nanowire and
the cavity used in section IV. Appendix B gives details on the calculation of
the nanowire conductance. Appendix C discusses the behavior of the cavity in
the classical regime, i.e. when a large number of photons are present in the cavity.

\section{Low-energy Hamiltonian model of the four Majorana nanowire}

We consider a single channel nanowire subject to a Zeeman splitting $E_{z}$
and an effective gap $\Delta$ induced by a superconducting contact (Fig.1.a).
The nanowire presents a strong Rashba spin-orbit coupling with a
characteristic speed $\alpha_{so}$. The chemical potential $\mu$ in the
nanowire can be tuned locally by using electrostatic gates. The details of the
model are given in appendix A. For brevity, in this section, we discuss only
the main features of the model which lead us to the effective low energy
Hamiltonian used in the main text (Eqs.(\ref{low}) and (\ref{coupl})). We note
$\mu_{c}=\sqrt{E_{z}^{2}-\Delta^{2}}$ the chemical potential below which the
wire is in a topological phase\cite{Lutchyn,Oreg}. The wire has two
topological regions $\mu=\mu_{1}<\mu_{c}$ with length $L_{T}$ surrounded by
three non topological regions $\mu=\mu_{0}>\mu_{c}$, with $L_{NT}$ the length
of the central non topological region (Fig.1.b). MBSs appear in the nanowire
at the interfaces between topological and non topological phases, for
coordinates $z\simeq z_{i}$ with $i\in\{1,2,3,4\}$, $z_{1}=0$, $z_{2}=L_{T}$,
$z_{3}=L_{T}+L_{NT}$, and $z_{4}=L_{NT}+2L_{T}$. In the topological phases,
the wavefunction corresponding to MBS $i$ decays exponentially away from
$z=z_{i}$ with the characteristic vector $k_{m}(\mu_{1})=(\Delta-\sqrt
{E_{z}^{2}-\mu_{1}^{2}})/\hbar\alpha_{so}<0$ (see Appendix A for details). In
the non-topological phases, the decay of the MBSs is set by the two
characteristic vectors $k_{p/m}(\mu_{0})=(\Delta\pm\sqrt{E_{z}^{2}-\mu_{0}%
^{2}})/\hbar\alpha_{so}>0$. The difference in the number of characteristic
vectors from the topological to the non-topological phases is fundamentally
related to the existence of the topological phase transition in the nanowire.
Away from the topological transition, one can introduce a Majorana fermionic
operator $\gamma_{i}$ such that $\gamma_{i}^{\dag}=\gamma_{i}$ and $\gamma
_{i}^{2}=1/2$ to describe MBS $i$.

In a real system, due to the finite values of $L_{T}$ and $L_{NT}$, the
different MBSs overlap. The resulting coupling can be described with the low
energy Hamiltonian:%
\begin{equation}
H_{wire}=2i\epsilon(\gamma_{1}\gamma_{2}+\gamma_{3}\gamma_{4})+2i\widetilde
{\epsilon}\gamma_{2}\gamma_{3}\label{low}%
\end{equation}
with $\epsilon\simeq\lambda_{\epsilon}e^{k_{m}(\mu_{1})L_{T}}$ and
$\widetilde{\epsilon}\simeq\lambda_{\widetilde{\epsilon}}e^{-k_{m}(\mu
_{0})L_{NT}}$. Note that $\epsilon$ and $\widetilde{\epsilon}$ are purely real
because the Majorana operators are self-adjoint and $H$ must be Hermitian.
\ The coefficients $\lambda_{\epsilon}$ and $\lambda_{\widetilde{\epsilon}}$
depend on $\mu_{0}$, $\mu_{1}$, $E_{z}$ and $\Delta$ (see Appendix A.5).
Importantly, the coupling energies $\epsilon$ and $\widetilde{\epsilon}$
depend exponentially on $L_{T}$ and $L_{NT}$, as a direct consequence from the
exponentially localized nature of MBSs. Furthermore, the vectors $k_{m}%
(\mu_{1})$ and $k_{m}(\mu_{0})$ vanish for $\mu_{1}=\mu_{c}$ and $\mu_{0}%
=\mu_{c}$, respectively, or in other terms the spatial extension of the MBSs
increases when one approaches the topological transition. In this limit, large
values of $\epsilon$ and $\widetilde{\epsilon}$ can be obtained. However, it
should be noted that the use of Eq.(\ref{low}) is justified provided the
nanowire is operated far enough from the topological transition. We have
checked that this is the case for the parameters used in Figs. 2 and 5. This
point will be discussed in more details in section VI.

\section{Tunnel spectroscopy of the nanowire}

\begin{figure}[h]
\includegraphics[width=0.6\linewidth]{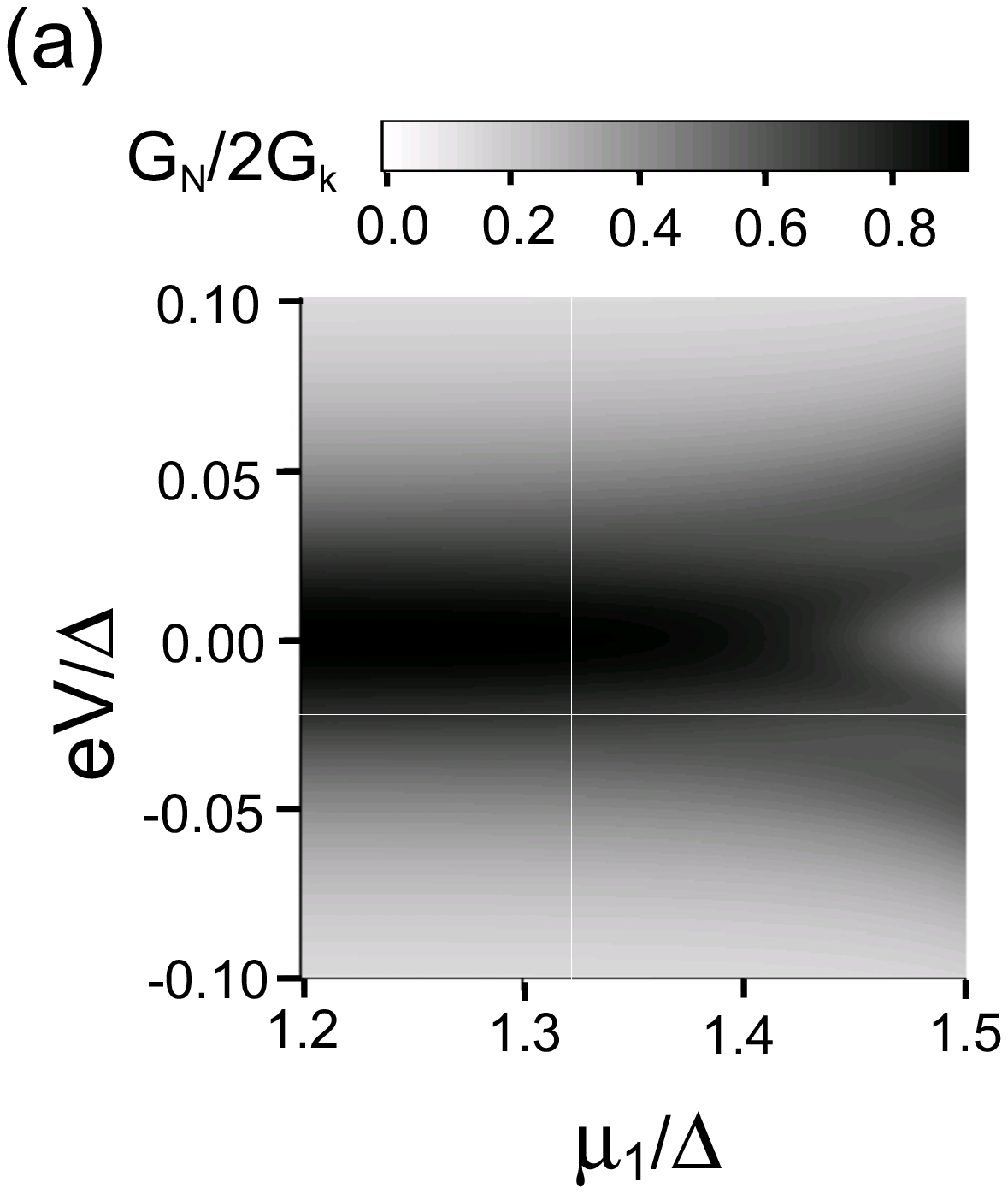}\newline%
\includegraphics[width=0.65\linewidth]{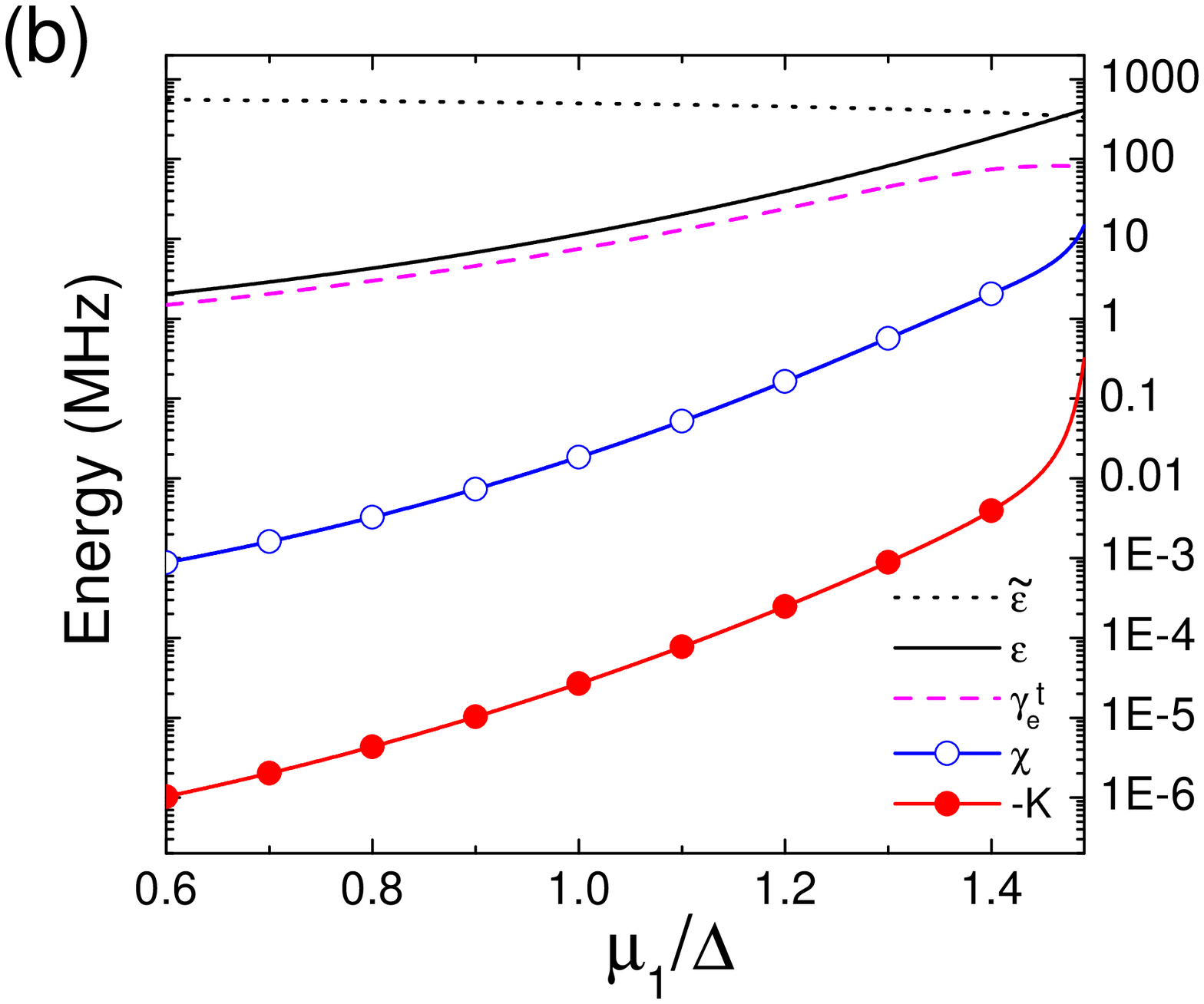}\caption{a. Conductance
$G_{N}$ between the normal metal contact and the ground, versus the bias
voltage $V$ and the chemical potential $\mu_{1}$. b. Coupling energies
$\varepsilon$ and $\tilde{\varepsilon}$ (full and dotted black lines),
transverse coupling $\gamma_{e}^{t}$ (pink dashed line), dispersive shift
$\chi$ of the cavity frequency (blue line with open circles), and Kerr
nonlinearity $K$ (red line with full circles) versus $\mu_{1}$. In this
figure, we have used $\Delta=500~\mathrm{\mu eV}$, $E_{z}/\Delta=2$, $\mu
_{0}/\Delta=1.9$, $\alpha_{so}\sim8.10^{4}m.s^{-1}$, $L_{T}=L_{NT}%
=1000~\mathrm{nm}$, $\alpha_{c}V_{rms}=4~\mathrm{\mu V}$, $\omega_{cav}%
/2\pi=8~\mathrm{GHz}$, $\Gamma=2~\mathrm{\mu eV}$, $T=10~\mathrm{mK}$ and
$G_{k}=e^{2}/h$. The topological transition is located well outside the
$\mu_{1}$ range considered here since $\mu_{c}/\Delta=1.73$. }%
\end{figure}

The simplest idea to probe MBSs is to perform a tunnel spectroscopy of the
nanowire by placing a normal metal contact biased with a voltage $V$ on the
nanowire, close to MBS 1 for instance (Fig.1.a). A current can flow between
the normal metal contact and the ground, through the MBSs and the grounded
superconducting contact shown in Fig.1a., which is tunnel coupled to the
nanowire. To describe the main properties of the conductance $G_{N}$ between
the normal metal contact and the ground, it is sufficient to assume an energy
independent tunnel rate $\Gamma$ between MBS 1 and the contact. The details of
the calculation are presented in appendix B. Figure 2.a shows $G_{N}$ as a
function of $\mu_{1}$ and $V$, for realistic parameters (see legend of Fig.2).
For $\mu_{0}$ and $\mu_{1}$ relatively close to $\mu_{c}$, $\epsilon$ and
$\widetilde{\epsilon}$ can be comparable or larger than $\Gamma$ and the
temperature scale $k_{B}T$. Hence, four conductance peaks appear at voltages
corresponding to the eigenenergies $(\pm\hbar\omega_{e}\pm\hbar\omega_{o})/2$
of $H_{wire}$, with $\hbar\omega_{e}=2\sqrt{4\epsilon^{2}+\widetilde{\epsilon
}^{2}}$ and $\hbar\omega_{o}=2\widetilde{\epsilon}$. In this regime, the
current flows through the four MBSs which are coupled together, as represented
in Fig.3.a. As $\mu_{1}$ decreases, the coupling between MBS 1 and the other
MBSs\begin{figure}[h]
\includegraphics[width=0.65\linewidth]{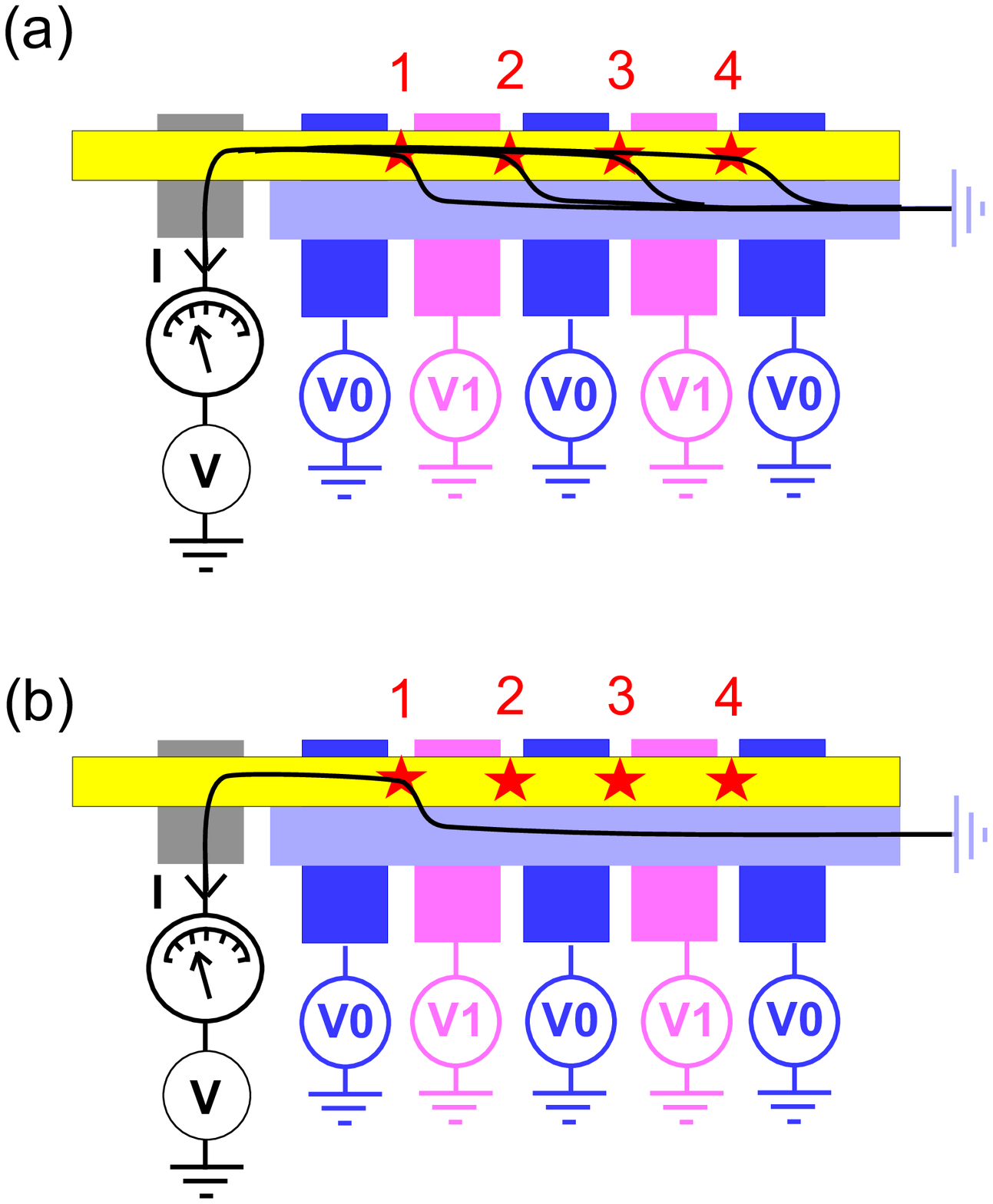}\newline\caption{Schematic
representation of the current flow in the nanocircuit depending on the value
of $\mu_{1}$. Panel (a) corresponds to $\mu_{0}$ and $\mu_{1}$ close to
$\mu_{c}$ so that $\varepsilon$ and $\widetilde{\varepsilon}$ are finite and
the four MBSs are coupled together. In this regime the current flows from the
superconducting contact (purple) to the normal metal contact (grey) through
the four Majorana bound states. Panel (b) corresponds to a value of $\mu_{1}$
far below $\mu_{c}$ so that the coupling $\varepsilon$ between MBS1 and other
MBSs vanishes. \ In this case, the current flows from the superconducting
contact to the normal metal contact through MBS 1 only.}%
\end{figure}\ disappear ($\epsilon\rightarrow0$), so that there remains only a
zero energy conductance peak which is due to transport through MBS 1, as
represented in Fig.3.b. Similar features can be caused by other effects such
as weak antilocalization, Andreev resonances or a Kondo
effect\cite{Lee,Liu,Pikulin,Rainis,Zareyan,CottetSF}. It is therefore
important to search for other ways to probe MBSs more specifically. We show in
the following that coupling the nanowire to a photonic cavity can give direct
signatures of the self-adjoint character of MBSs and their exponential
confinement. In the rest of the paper, we omit the explicit description of the
normal metal contact. The Majorana system could be affected by decoherence,
due to the normal metal contact or background charge fluctuators in the
vicinity of the nanowire, for instance. However the detection scheme we
present below is to a great extent immune to decoherence because it leaves the
Majorana system in its ground state (we use $\hbar\omega_{e/o}\gg k_{B}T$).

\section{Coupling between the nanowire and a microwave cavity}

We assume that the nanowire is placed between the center and ground conductors
of a coplanar waveguide cavity (Fig.1.a). We take into account a single mode
of the cavity, corresponding to a photon creation operator $a^{\dag}$. There
exists a capacitive coupling between the nanowire and the cavity, which is
currently observed in experiments\cite{Delbecq,DQD,Petersson}. More precisely,
the nanowire chemical potential is shifted by $\mu_{ac}=e\alpha_{c}%
V_{rms}(a+a^{\dag})$, with $V_{rms}$ the rms value of the cavity vacuum
voltage fluctuations and $\alpha_{c}$ a capacitive ratio. This leads to the
system Hamiltonian
\begin{equation}
H=H_{wire}+h_{int}(a+a^{\dag})+\hbar\omega_{cav}a^{\dag}a\label{coupl}%
\end{equation}
with $h_{int}=2i\beta(\gamma_{1}\gamma_{2}+\gamma_{3}\gamma_{4})+2i\widetilde
{\beta}\gamma_{2}\gamma_{3}$, $\beta\simeq\lambda_{\beta}(L_{T}/l_{c}%
)\epsilon$, $\widetilde{\beta}\simeq\lambda_{\widetilde{\beta}}(L_{NT}%
/l_{c})\widetilde{\epsilon}$ and $l_{c}=\hbar\alpha_{so}/e\alpha_{c}V_{rms}$.
Note that $\beta$ and $\widetilde{\beta}$ are purely real, due again to
$\gamma_{i}^{\dag}=\gamma_{i}$. The coefficients $\lambda_{\beta}$ and
$\lambda_{\widetilde{\beta}}$ depend on $\mu_{0(1)}$ and $E_{z}$ (see appendix
A.5). The term in $h_{int}$ is caused by the potential shift $\mu_{ac}$. Due
to $h_{int}$, cavity photons modify the coupling between MBSs, as represented
schematically in Fig.4. \begin{figure}[ptb]
\includegraphics[width=1.\linewidth]{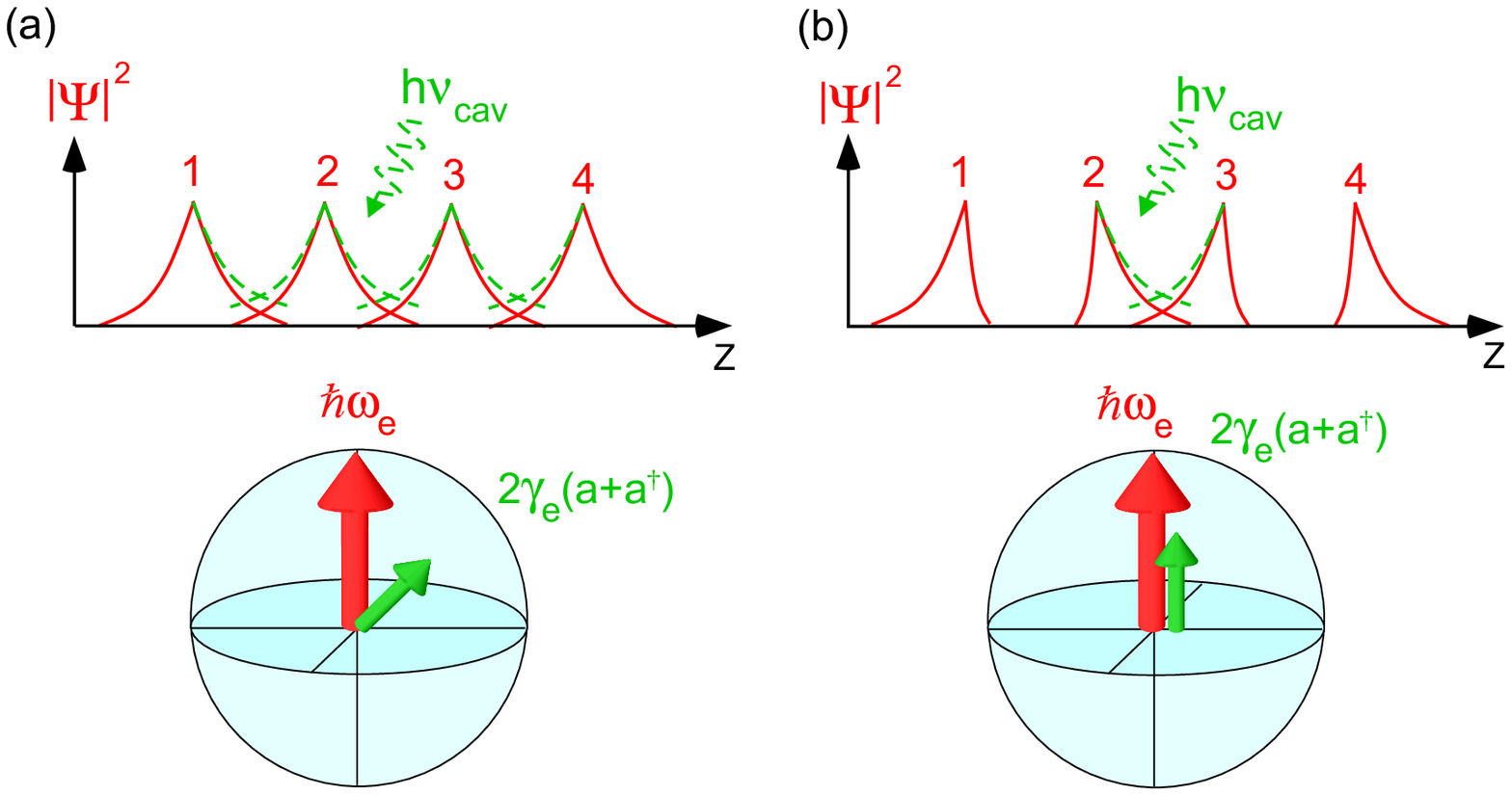}\caption{Schematic
representation of the coupling mechanism between MBSs and cavity photons a.
When $\mu_{0}$ and $\mu_{1}$ are close to the critical potential $\mu_{c}$,
cavity photons modify the coupling between all consecutive MBSs. This yields a
coupling between the nanowire and the cavity with a transverse component in
the MBSs even charge sector, represented here on a Bloch sphere. b. When
$\mu_{1}$ is far below $\mu_{c}$, only MBSs 2 and 3 remain coupled by cavity
photons. In this case one can only have a longitudinal coupling in the
nanowire even charge sector, as a direct consequence of the self-adjoint
character of Majorana fermion operators. }%
\end{figure}Remarkably, $h_{int}$ has a form similar to $H_{wire}$, with
coefficients $\beta$ and $\widetilde{\beta}$ containing the same exponential
dependence on $L_{T}$ and $L_{NT}$ as $\epsilon$ and $\widetilde{\epsilon}$,
because $\mu_{ac}$ is spatially constant along the nanowire. Hence, the
amplitude of $h_{int}$ directly depends on the MBSs exponential overlaps.

One can reveal important properties of MBSs by varying $\mu_{1}$, with
$\mu_{0}$ constant. Let us assume that $\mu_{0}$ is relatively close to
$\mu_{c}$ so that $\widetilde{\epsilon}$ and $\widetilde{\beta}$ can be
considered as finite. When $\mu_{1}$ is also close to $\mu_{c}$, $\epsilon$
and $\ \beta$ are finite, and in general $\beta\widetilde{\epsilon}%
\neq\widetilde{\beta}\epsilon$, so that $h_{int}$ and $H_{wire}$ are not
proportional. This enables the existence of a transverse coupling between the
Majorana system and the cavity, i.e. the cavity photons can induce changes in
the state of the Majorana system, as we will see in more details below. In
contrast, for $\mu_{1}$ far below $\mu_{c}$, $\epsilon$ and $\beta$ vanish
because MBSs are strongly localized in the topological phases. This means that
MBSs 2 and 3 remain coupled together and they are also coupled to the cavity,
but MBSs 1 and 4 become isolated and thus irrelevant for the cavity (Fig 4.b).
In this limit, $H$ takes the form of the Hamiltonian of a single pair of
coupled Majorana fermions, i.e.
\begin{equation}
H^{\prime}\simeq2i\widetilde{\epsilon}\gamma_{2}\gamma_{3}+2i\widetilde{\beta
}\gamma_{2}\gamma_{3}(a+a^{\dag})\label{b0}%
\end{equation}
Note that the eigenvalues of $H^{\prime}$ have a twofold degeneracy due to the
existence of the isolated MBSs 1 and 4. Both terms in the Hamiltonian
(\ref{b0}) have the same structure, or in other terms $h_{int}$ and $H_{wire}$
are proportional, due to constraints imposed by the self-adjoint character of
Majorana fermions. Indeed, a quadratic Hamiltonian involving only MBSs 2 and 3
must necessarily be proportional to $i\gamma_{2}\gamma_{3}$ since the terms
$\gamma_{2}^{\dag}\gamma_{2}$ and $\gamma_{3}^{\dag}\gamma_{3}$ are
proportional to the identity and therefore inoperant for self-adjoint
fermions. As a result, the coupling between the cavity and the Majorana system
becomes purely longitudinal, as discussed in more details below.

To discuss more precisely the structure of the coupling between the nanowire
and the cavity, it is convenient to reexpress $H$ in terms of ordinary
fermionic operators. One possibility is to use the two fermions $c_{L}^{\dag
}=(\gamma_{1}-i\gamma_{2})/\sqrt{2}$ and $c_{R}^{\dag}=(\gamma_{3}-i\gamma
_{4})/\sqrt{2}$. A second possibility is to use $c_{m}^{\dag}=(\gamma
_{2}-i\gamma_{3})/\sqrt{2}$ \ and $c_{e}^{\dag}=(\gamma_{1}-i\gamma_{4}%
)/\sqrt{2}$. Depending on the cases, it is more convenient to use the first or
the second possibility. We also define the occupation numbers $n_{f}%
=c_{f}^{\dag}c_{f}$, for $f\in\{L,R,e,m\}$. In the discussion following, we
recover the fact that in a closed system made of several Majorana bound
states, the parity of the total number of fermions is conserved\cite{reviews}.
Note that in our system, the total fermions numbers $N_{tot}=n_{L}+n_{R}$ or
$N_{tot}^{\prime}=n_{e}+n_{m}$ are not equivalent since they do not commute,
but their parity $P=-4\gamma_{1}\gamma_{2}\gamma_{3}\gamma_{4}$ is the same.

For $\mu_{1}$ far below $\mu_{c}$, it is convenient to use the basis of the
fermions $e$ and $m$ to reexpress the Hamiltonian as%
\begin{equation}
H^{\prime}=(\widetilde{\epsilon}+\widetilde{\beta}(a+a^{\dag}))(2n_{m}%
-1)+\hbar\omega_{cav}a^{\dag}a\label{b1}%
\end{equation}
One can note that $c_{e}^{\dag}$ and $c_{e}$ do not occur in $H^{\prime}$,
therefore the $e$ fermionic degree of freedom can be disregarded. Moreover,
one has $[H,n_{m}]=0$, which means that the number of fermions of type $m$ (or
equivalently the parity of $n_{m}$) is a conserved quantity, as expected for
an (effective) system of 2 Majorana bound states. Hence, the coupling to the
cavity cannot change $n_{m}$, or in other terms it cannot affect the state of
the Majorana fermions. This means that in this limit, the coupling between the
nanowire and the cavity can only be longitudinal as already mentioned above.

When $\mu_{0}$ and $\mu_{1}$ are both close enough to $\mu_{c}$, it is more
convenient to use the basis of fermions $L$ and $R$. We define
$(0,0)=\left\vert 0\right\rangle $, $(1,0)=c_{L}^{\dag}\left\vert
0\right\rangle $, $(0,1)=c_{R}^{\dag}\left\vert 0\right\rangle $, and
$(1,1)=c_{L}^{\dag}c_{R}^{\dag}\left\vert 0\right\rangle $. Since $\epsilon$,
$\beta$, $\widetilde{\epsilon}$ and $\widetilde{\beta}$ are finite, we have a
fully effective four-Majorana system whose Hamiltonian writes:%

\begin{align}
H &  =2\left(  \epsilon+\beta(a+a^{\dag})\right)  (n_{L}+n_{R}-1)\label{bib}\\
&  +\left(  \widetilde{\epsilon}+\widetilde{\beta}(a+a^{\dag})\right)  \left(
c_{L}^{\dag}c_{R}-c_{L}c_{R}^{\dag}+c_{L}^{\dag}c_{R}^{\dag}-c_{L}%
c_{R}\right)  \nonumber\\
&  +\hbar\omega_{cav}a^{\dag}a\nonumber
\end{align}
One can check from this equation that the parity of $N_{tot}=n_{L}+n_{R}$ is
conserved as expected. However, since we have now 2 fermionic degrees of
freedom fully involved in the Hamiltonian, we have to consider the two parity
subspaces $\mathcal{E}_{e}=\{(0,0),(1,1)\}$ and $\mathcal{E}_{o}%
=\{(0,1),(1,0)\}$, each with a dimension 2. The conservation of the total
fermion parity forbids transitions between $\mathcal{E}_{e}$ and
$\mathcal{E}_{o}$, as can be checked from the structure of Eq. (\ref{bib}).
However, nothing forbids the cavity to induce transitions inside each of the
parity subspaces, as shown by the structure of the term in $\widetilde{\beta}%
$. Therefore, when the 4 Majorana states are effective, a transverse coupling
between the nanowire and the cavity is possible.

To push further our analysis, it is convenient to introduce effective spin
operators $\overrightarrow{\sigma}_{e}=\{\sigma_{e,X},\sigma_{e,Z}\}$ and
$\overrightarrow{\sigma}_{o}=\{\sigma_{o,X},\sigma_{o,Z}\}$ operating in the
subspaces $\mathcal{E}_{e}$ and $\mathcal{E}_{o}$ respectively, i.e.
$\sigma_{e,z}=1-c_{L}^{\dag}c_{L}-c_{R}^{\dag}c_{R}$, $\sigma_{e,x}%
=c_{L}^{\dag}c_{R}^{\dag}-c_{L}c_{R}$, $\sigma_{o,z}=c_{L}^{\dag}c_{L}%
-c_{R}^{\dag}c_{R}$, and $\sigma_{o,x}=(c_{L}^{\dag}c_{R}-c_{L}c_{R}^{\dag}).$
For convenience we rotate the spin operators as $\tilde{\sigma}_{e,z}%
=(-2\epsilon\sigma_{e,z}-\widetilde{\epsilon}\sigma_{e,x})/\sqrt{4\epsilon
^{2}+\widetilde{\epsilon}^{2}}$, $\tilde{\sigma}_{e,x}=(-\widetilde{\epsilon
}\sigma_{e,z}+2\epsilon\sigma_{e,x})/\sqrt{4\epsilon^{2}+\widetilde{\epsilon
}^{2}}$ and $\tilde{\sigma}_{o,z}=-\sigma_{o,x}$. We finally obtain
\begin{equation}
H_{wire}=(\hbar\omega_{e}\tilde{\sigma}_{e,z}+\hbar\omega_{o}\tilde{\sigma
}_{o,z})/2
\end{equation}
and
\begin{equation}
h_{int}=\gamma_{e}^{t}\tilde{\sigma}_{e,x}+\gamma_{e}^{l}\tilde{\sigma}%
_{e,z}+\gamma_{o}^{l}\tilde{\sigma}_{o,z}%
\end{equation}
with
\begin{equation}
\gamma_{o}^{l}=\widetilde{\beta}%
\end{equation}%
\begin{equation}
\gamma_{e}^{l}=(4\beta\epsilon+\widetilde{\beta}\widetilde{\epsilon}%
)/\sqrt{4\epsilon^{2}+\widetilde{\epsilon}^{2}}%
\end{equation}
and%
\begin{equation}
\gamma_{e}^{t}=\frac{2\left(  \beta\widetilde{\epsilon}-\widetilde{\beta
}\epsilon\right)  \left(  \sqrt{4\epsilon^{2}+\widetilde{\epsilon}^{2}%
}-2\epsilon\right)  }{\sqrt{32\epsilon^{4}+12\epsilon^{2}\widetilde{\epsilon
}^{2}+\widetilde{\epsilon}^{4}-4\epsilon(4\epsilon^{2}+\widetilde{\epsilon
}^{2})^{3/2}}}%
\end{equation}
These expressions show that the cavity couples longitudinally to the odd
charge sector, whereas the coupling to the even charge sector can have a
transverse component $\gamma_{e}^{t}$ because $\beta\widetilde{\epsilon}%
\neq\widetilde{\beta}\epsilon$ in general (Fig 4.a). The absence of transverse
coupling in the odd charge sector is a consequence of the particular
symmetries that we have assumed in our system, as will be discussed in section
VI. For $\mu_{1}$ far below $\mu_{c}$, $\epsilon$ and $\beta$ vanish thus
$H^{\prime}\simeq%
{\textstyle\sum\nolimits_{j\in\{e,o\}}}
H_{j}$ with
\begin{equation}
H_{j}=\frac{\hbar\omega_{j}}{2}\tilde{\sigma}_{j,z}+\gamma_{j}^{l}(a+a^{\dag
})\tilde{\sigma}_{j,z} \label{b2}%
\end{equation}
Both terms in the expression (\ref{b2}) have the same structure in the
effective spin space. Thus, we recover again the fact that the coupling
between the Majorana system and the cavity becomes purely longitudinal for
$\mu_{1}$ far below $\mu_{c}$. The cancellation of the transverse coupling
between the nanowire and the cavity is fundamentally related to the
self-adjoint character of MBSs which imposes the forms (\ref{b0}), or
equivalently (\ref{b1}) or (\ref{b2}) in the case of a 2 Majorana system.

In conclusion, one can reveal important properties of MBSs by varying $\mu
_{1}$, with $\mu_{0}$ constant. The vanishing of $\gamma_{e}^{t}$ for $\mu
_{1}$ far below $\mu_{c}$ in spite of the fact that $\widetilde{\epsilon}$ and
$\widetilde{\beta}$ remain finite represents a strong signature of the
self-adjoint character of MBSs. In addition, probing the dependence of
$\gamma_{e}^{t}$ on $\mu_{1}$ could reveal the exponential confinement of MBSs
since for $\mu_{1}$ sufficiently below $\mu_{c}$, $\gamma_{e}^{t}\simeq
4(\beta\widetilde{\epsilon}-\widetilde{\beta}\epsilon)/\widetilde{\epsilon}$
scales with $e^{k_{m}(\mu_{1})L_{T}}$. Also note that, in principle, for
$\mu_{0}$ and $\mu_{1}$ close enough to $\mu_{c}$, $\gamma_{e}^{t}$ can be
large due to the large spatial extension of MBSs (see Fig.2.b). To test these
properties, it is important to have an experimental access to $\gamma_{e}^{t}%
$. We show below that this is feasible due to the strong effects of
$\gamma_{e}^{t}$ on the cavity dynamics.

\section{Behavior of the microwave cavity in the dispersive regime}

In the dispersive (i.e. non resonant) regime, the transverse coupling
$\gamma_{e}^{t}$ between the effective spin $\overrightarrow{\tilde{\sigma}%
}_{e}$ and the cavity allows for fast high order processes in which the
population of the effective spin is changed virtually. This effect can be
described by using an adiabatic elimination followed by a projection on the
nanowire ground state\cite{Cohen-Tannoudji}. This yields an effective cavity
Hamiltonian:%
\begin{equation}
H_{adiab}=\hbar\omega_{cav}a^{\dag}a+\chi a^{\dag}a+K(a^{\dag})^{2}%
a^{2}+o(\gamma_{e}^{6})
\end{equation}
with $\omega_{cav}$ the cavity frequency,%
\begin{equation}
\chi=\left(  2(\gamma_{e}^{t})^{2}\omega_{e}/(\omega_{cav}^{2}-\omega_{e}%
^{2})\right)  +o(\gamma_{e}^{4})
\end{equation}%
\begin{align}
K &  =d^{-1}(\gamma_{e}^{t})^{2}(\gamma_{e}^{l})^{2}\omega_{e}(8\omega
_{cav}^{4}+20\omega_{e}^{4}-28\omega_{e}^{2}\omega_{cav}^{2})\label{K}\\
&  +d^{-1}(\gamma_{e}^{t})^{4}\omega_{e}(8\omega_{cav}^{4}-6\omega_{e}%
^{4}+22\omega_{e}^{2}\omega_{cav}^{2})+o(\gamma_{e}^{6})\nonumber
\end{align}
and $d=(\omega_{e}^{2}-\omega_{cav}^{2})^{3}(4\omega_{cav}^{2}-\omega_{e}%
^{2})$. The transverse coupling $\gamma_{e}^{t}$ causes a cavity frequency
shift $\chi$ and a non-linear term proportional to $K$, similar to the Kerr
term widely used in nonlinear optics. Figure 2.b illustrates that $\epsilon$,
$\gamma_{e}^{t}$, $\chi$ and $K$ quickly vanish when $\mu_{1}$ goes far below
$\mu_{c}$. In this limit, $\chi$ and $K$ both scale with $(\gamma_{e}^{t}%
)^{2}$ because due to $\gamma_{e}^{t}\ll\gamma_{e}^{l}\sim\widetilde{\beta}$,
the first contribution in Eq.(\ref{K}) dominates $K$. For the realistic
parameters used in this figure, $\chi$ varies from about $14~$\textrm{MHz} to
$9~10^{-4}~$\textrm{MHz}. In practice, $\chi$ can be measured
straightforwardly by measuring the response of the cavity to an input signal
with a small power, for values down to $-10^{-3}~$\textrm{MHz} at least. The
upper value $\chi\simeq14~\mathrm{MHz}$ is comparable to what has been
obtained with strongly coherent two level systems slightly off-resonant with a
microwave cavity\cite{Majer}. Having a significant Kerr nonlinearity is more
specific to the ultra-strong spin/cavity coupling regime, which we obtain in
our system because MBSs have a large spatial extension near the topological
transition. In Fig.2.b, the Kerr constant $K$ varies from $-0.31~$\textrm{MHz}
to $-10^{-6}~$\textrm{MHz}. The value $K=-0.31~$\textrm{MHz} is comparable to
nonlinearities obtained recently with microwave resonators coupled to
Josephson junctions\cite{Ong,Kirchmair}. However, it is important to notice
that our $\chi$ and $K$ term have an approximate exponential dependence on
$\mu_{1}$ due to the factor $e^{k_{m}(\mu_{1})L_{T}}$ appearing in $\gamma
_{e}^{t}$, which is very specific to MBSs.

Figure 5 illustrates how to measure $K$ by probing the response of the cavity
to an input microwave signal. We note $\gamma_{in/out}$ the photonic coupling
rate between the input/output port and the cavity, and $\gamma$ the total
decoherence rate of cavity photons. If $K$ is small, it can be revealed by
applying to the cavity a steady signal which drives the resonator into a
semi-classical regime \cite{Ong} (see details in Appendix C). The
semiclassical response of the cavity to a forward and backward sweep of
$\omega_{RF}$ becomes hysteretic for a critical power $P_{in}^{c}=4\gamma
p_{in}^{0}/3\sqrt{3}\left\vert K\right\vert $ which can be used to determine
$\left\vert K\right\vert $, with $p_{in}^{0}=\hbar\omega_{cav}\gamma
^{2}/2\gamma_{in}$ the single photon input power\cite{Yurke} (Fig.5.a). Such a
technique should allow one to observe MBSs relatively far from the topological
transition, by using a high input power which compensates for the smallness of
$K$. For the measurement of $\chi$, one does not benefit from such an
advantage, hence we believe that the measurement of $K$ can enable one to
follow the behavior of MBSs on a wider range of $\mu_{1}$. For the highest
values of $K$, the classically-defined critical power $P_{in}^{c}$ is so small
that the resonator is still in a quantum regime at this power. In this case
one can directly observe the cavity nonlinearity with a low input power, by
performing a tomographic measurement of the cavity Husimi Q-function
$Q(\alpha)=Tr[\rho_{cav}(t)\left\vert \alpha\right\rangle \left\langle
\alpha\right\vert ]$ at a time $\Delta t$ after switching off the input
bias\cite{Kirchmair} (Fig.5.b). Here $\rho_{cav}(t)$ is the cavity density
matrix, $\left\vert \alpha\right\rangle =e^{-\left\vert \alpha\right\vert
^{2}/2}%
{\textstyle\sum\nolimits_{n}}
\alpha^{n}\left\vert n\right\rangle /\sqrt{n!}$ denotes a cavity coherent
state and $\left\vert n\right\rangle $ a cavity Fock state with $n$ photons.
The $K$ term can produce a strong photon amplitude squeezing which can be
calculated for $\hbar\omega_{cav}\gg k_{B}T$ following Ref.\cite{Milburn}%
.\begin{figure}[ptb]
\includegraphics[width=0.65\linewidth]{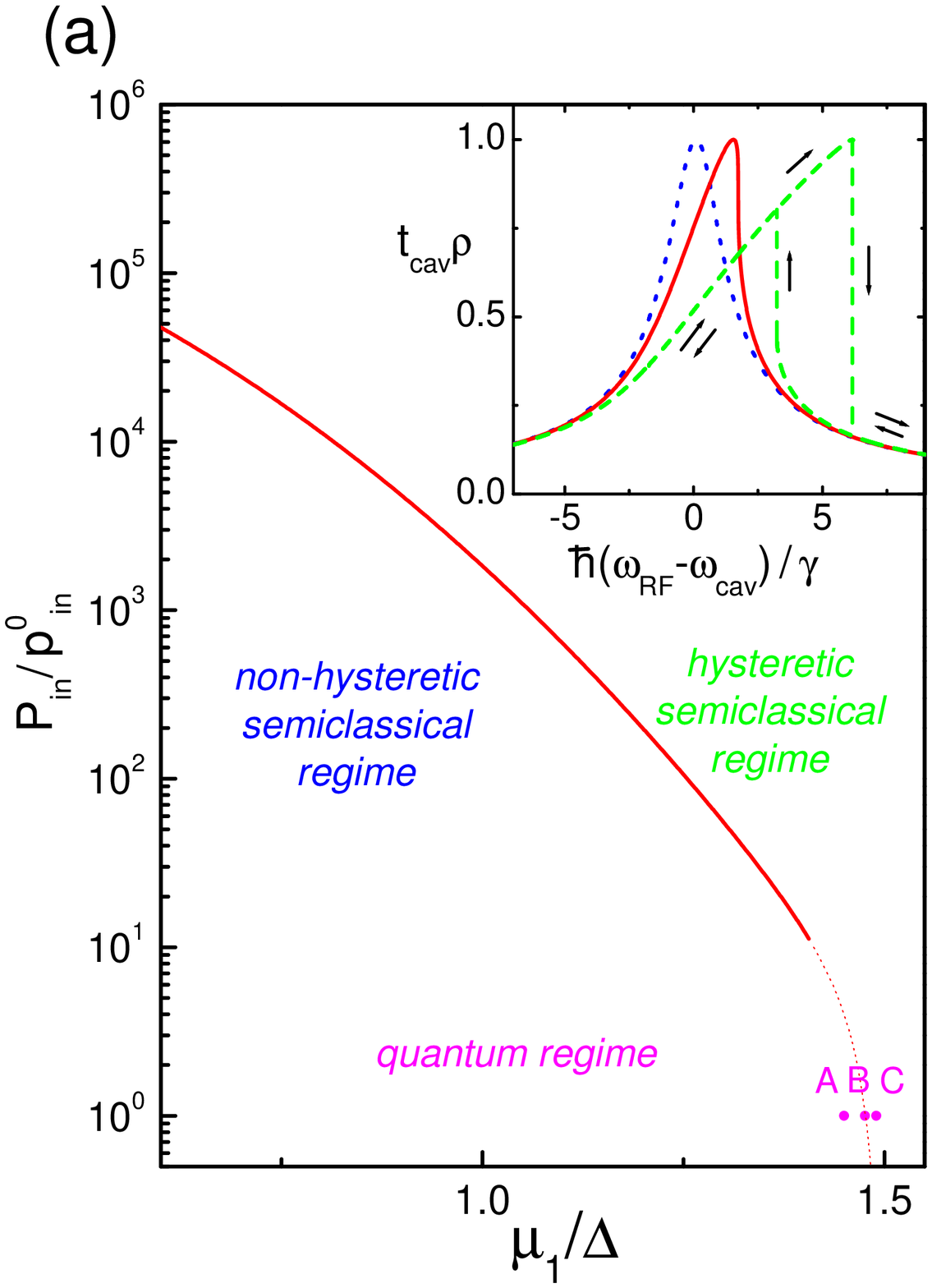}\newline%
\includegraphics[width=0.5\linewidth]{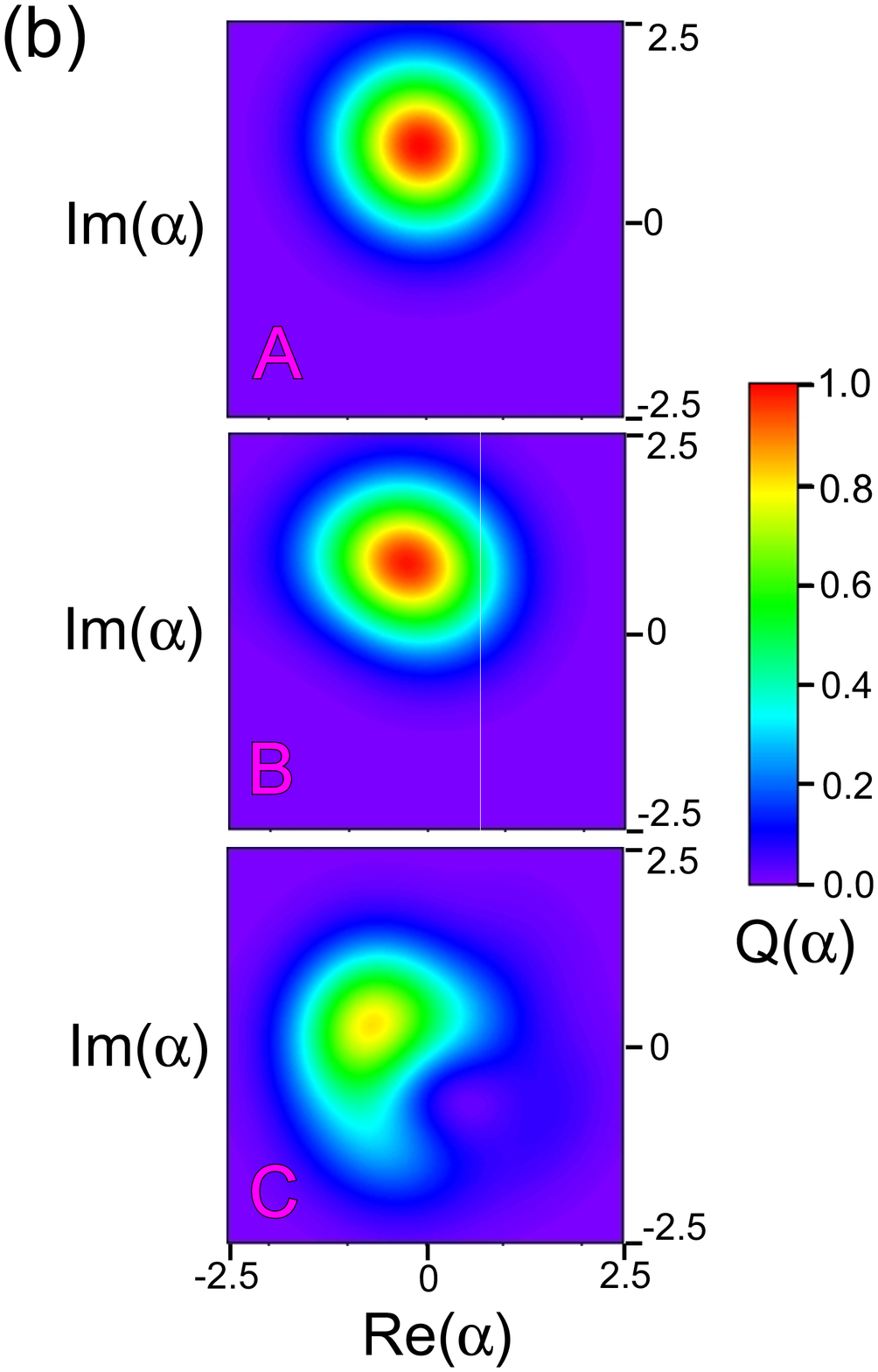}\caption{a. Diagram of the
cavity behavior depending on the steady power $P_{in}$ applied to its input
port and the nanowire potential $\mu_{1}$. For $P_{in}>P_{in}^{crit}$, the
behavior of the cavity becomes hysteretic. Inset: Modulus $t_{cav}$ of the
cavity transmission for a forward and backward sweep of $\omega_{RF}$, for
$P_{in}=P_{in}^{crit}/4$ (blue dotted lines), $P_{in}=P_{in}^{crit}$ (full red
lines) and $P_{in}=4P_{in}^{crit}$ (green dashed lines). b. Cavity Husimi
function $Q(\alpha)$ at points A, B and C of Fig.4.a, at a time $\Delta
t=175\mathrm{ns}$ after switching off an input power imposing a coherent
cavity state $\left\vert i\right\rangle $. We have used the same parameters as
in Fig.2, $\rho=\gamma/2\sqrt{\gamma_{in}\gamma_{out}}$ and $Q_{cav}%
=\hbar\omega_{cav}/2\pi\gamma=10000$.}%
\end{figure}

\section{Discussion}

Before concluding, we discuss various simplifications used in the description
of our results. First, we find that the nanowire odd charge sector does not
have a transverse coupling to the cavity due to the symmetry between the
sections 1-2 and 3-4 of the nanowire. If these sections had different lengths
or parameters, a coupling to the odd charge sector would be possible, but we
expect qualitatively similar results in this case because in the limit of
$\mu_{1}$ far below $\mu_{c}$, the self-adjoint character of Majorana
operators still imposes a system Hamiltonian of the form (\ref{b0}), or
equivalently (\ref{b1}) or (\ref{b2}), and the coupling between MBSs 1 and 2
(3 and 4) should still depend exponentially on $L_{T}$. Second, with our
nanowire model, a topological transition also occurs for $\mu=-\mu_{c}$.
Therefore, upon decreasing $\mu$ the absolute values of $\epsilon$, $\beta$,
$\gamma_{e}^{t}$, $\chi$ and $K$ reach minima for $\mu\sim0$, and increase
again when $\mu_{1}$ approaches $-\mu_{c}$. We have not discussed this limit
because it gives results similar to $\mu\rightarrow\mu_{c}$.

Note that the use of the low energy Hamiltonian description, i.e.
Eqs.(\ref{low}) and (\ref{coupl}), is justified provided the nanowire is
operated far enough from the topological transition. This is essential to have
large enough nanowire bandgaps. These bandgaps can be defined as $E_{b}%
^{1(0)}=(2\Delta^{2}+\mu_{1(0)}^{2}+\mu_{c}^{2}-2((\Delta^{2}+\mu_{1(0)}%
^{2})(\Delta^{2}+\mu_{c}^{2}))^{1/2})^{1/2}$ in the topological
(non-topological) sections of the nanowire. With the parameters range
considered in Figs. 2 and 4, one has $E_{b}^{1(0)}>35.6~\mathrm{GHz}$. In
comparison, our hybridized Majorana bound states lie at frequencies $\pm
\omega_{e}/2\pm\omega_{o}/2$ which lie in the interval $[-2.4~\mathrm{GHz}$,
$2.4~\mathrm{GHz}]$. Therefore, these bound states are well separated from the
continuum of states which exists above the nanowire gaps. With a typical
cavity ($\omega_{cav}=8~\mathrm{GHz}$), it is thus not possible to excite
quasiparticle above these gaps. Operating the device away from the topological
transition also grants that possible fluctuations of the nanowire potentials
due to charge fluctuators in the environment of the nanowire will not be
harmful. For the range of parameters considered in Figs. 2 and 4, one has
$\mid\mu_{0(1)}-\mu_{c}\mid>84~\mathrm{\mu eV}>eV_{ch}$, with $V_{ch}%
\simeq10~\mathrm{\mu V}$ the typical amplitude for charge noise in
semiconducting nanowires (see Ref. \cite{Petersson}). Charge noise is a low
frequency effect which should mainly smooth the measured $\chi$ and $K$ if one
stays away from the topological transition. This effect should not be dramatic
since we expect the exponential variation of $\chi$ and $K$ to occur on a wide
$\mu_{1}$ potential scale.

In more sophisticated models including disorder or several channels, the
occurrence of MBSs can be more complex (see e.g.
\cite{Pikulin,Liu,Brouwer,Potter,Lim}). Our setup precisely aims at testing
whether their exists regimes where the four-MBSs low energy description of
Eqs.(\ref{low}) and (\ref{coupl}) remains valid.\ In this limit, our findings
are very robust since they only rely on the fact that MBS have a self-adjoint
character and a gate-controlled spatial extension. Interestingly, a double
quantum dot (DQD) can also be coupled transversely to a microwave
cavity\cite{DQDth}, which leads to a cavity frequency shift, as confirmed by
recent experiments \cite{DQD,Petersson}. When the double dot and the cavity
are coupled dispersively, and the two dot orbitals resonant, the cavity
frequency shift and the DQD conductance are maximal. However, when the DQD
orbital energies or interdot hopping are varied to decrease the cavity
frequency shift, this also switches off the DQD conductance. In contrast, for
the system we consider here, the low energy conductance peak will persist in
spite of the decrease of $\chi$ and $K$. Hence, it can be useful to measure
simultaneously the cavity response and the nanosystem conductance to discard
spurious effects due to accidental quantum dots. Note that this does not make
our proposal more difficult to realize experimentally. Such joint measurements
are currently performed in experiments combining\ nanocircuits and coplanar
microwave cavities. This is a recent but mature technology as can been seen in
Refs. \cite{Delbecq,DQD,Petersson}.

\section{Conclusion}

In conclusion, we have considered a semiconducting nanowire device hosting
four MBSs coupled to a microwave cavity. This systems shows a cavity frequency
shift and a Kerr photonic nonlinearity when the nanowire is close enough to
the topological transition. These effects disappear when the nanowire gates
are tuned such that only two MBSs remain coupled to the cavity, due to the
self-adjoint character of MBSs which imposes strong constraints on the
cavity/nanowire coupling. Meanwhile, the low energy conductance peak caused by
the MBSs persists, a behavior which should be difficult to mimic with other
systems. The gate dependences of the cavity frequency shift and of the Kerr
nonlinearity should furthermore reveal the exponential confinement of MBSs.

\textit{We acknowledge discussions with G. Bastard, R. Feirrera, B. Huard, F.
Mallet, M. Mirrahimi and J.J. Viennot. This work was financed by the EU-FP7
project SE2ND[271554] and the ERC Starting grant CirQys.}

\section{Appendix A: One dimensional microscopic description of the
semiconducting nanowire}

\subsection{A.1 Initial one-dimensional Hamiltonian for the semiconducting
nanowire}

We describe the electronic dynamics in the nanowire with an effective
one-dimensional Hamiltonian%
\begin{equation}
\mathcal{H}_{1D}=%
{\textstyle\int}
dz[%
\begin{array}
[c]{cc}%
\Psi_{\uparrow}^{\dag}(z) & \Psi_{\downarrow}^{\dag}(z)
\end{array}
]H_{1D}\left[
\begin{array}
[c]{c}%
\Psi_{\uparrow}(z)\\
\Psi_{\downarrow}(z)
\end{array}
\right]
\end{equation}
with%
\begin{align}
H_{1D}(z) &  =-\frac{\hbar^{2}}{2m}\frac{\partial^{2}}{\partial z^{2}}%
+E_{z}\sigma_{z}-\mu(z)-\mu_{ac}\nonumber\\
&  -i\hbar(\alpha_{x}\sigma_{y}-\alpha_{y}\sigma_{x})\frac{\partial}{\partial
z}%
\end{align}
Here, $\Psi_{\sigma}^{\dag}(z)$ creates an electron with spin $\sigma$ at
coordinate $z$. An external magnetic field induces a Zeeman splitting $E_{z}$
in the nanowire. The chemical potential $\mu(z)$ can be controlled by using
electrostatic gates. The constants $\alpha_{x}$ and $\alpha_{y}$ account for
Rashba spin-orbit interactions corresponding to an effective electric field
which we express here in terms of a velocity vector $\overrightarrow
{\alpha_{so}}=\alpha_{x}\overrightarrow{u_{x}}+\alpha_{y}\overrightarrow
{u_{y}}$. The vector $\overrightarrow{\alpha_{so}}$ is expected to be
perpendicular to the nanowire\cite{Nadj}. Such a model is suitable provided
the description of the nanowire can be reduced to the lowest transverse
channel\cite{Lutchyn2}. We describe the coupling between the nanowire and the
cavity by using a potential term%
\begin{equation}
\mu_{ac}=e\alpha_{c}V_{rms}(a+a^{\dag})
\end{equation}
with $V_{rms}$ the rms value of the cavity vacuum voltage fluctuations and
$\alpha_{c}$ a dimensionless constant which depends on the values of the
different capacitances in the circuit. This type of coupling between a
nanoconductor and a cavity has been observed
experimentally\cite{Delbecq,DQD,Petersson}. In recent experiments, $\alpha
_{c}\sim0.3$ has been measured\cite{Delbecq}. Optimization of the microwave
designs could be used to increase this value.

\subsection{A.2 Bogoliubov-De Gennes equations for the nanowire}

One can describe the superconducting proximity effect inside the nanowire by
using%
\begin{equation}
\mathcal{H}_{BCS}=\mathcal{H}_{1D}+%
{\textstyle\int}
dz(\Delta\Psi_{\uparrow}^{\dag}(z)\Psi_{\downarrow}^{\dag}(z)+\Delta^{\ast
}\Psi_{\downarrow}(z)\Psi_{\uparrow}(z))
\end{equation}
with $\Delta$ a proximity-induced gap. We perform a Bogoliubov-De Gennes
transformation
\begin{align}
\gamma_{n}^{\dagger}  &  =%
{\textstyle\int}
dz^{\prime}(u_{\uparrow}(z^{\prime})\Psi_{\uparrow}^{\dag}(z^{\prime
})+u_{\downarrow}(z^{\prime})\Psi_{\downarrow}^{\dag}(z^{\prime})\nonumber\\
&  +v_{\uparrow}(z^{\prime})\Psi_{\uparrow}(z^{\prime})+v_{\downarrow
}(z^{\prime})\Psi_{\downarrow}(z^{\prime}))
\end{align}
such that $\mathcal{H}_{BCS}=\sum_{n}E_{n}\gamma_{n}^{\dagger}\gamma_{n}$. The
coefficients $u_{\uparrow}$, $u_{\downarrow}$, $v_{\uparrow}$ and
$v_{\downarrow}$ can be obtained by solving
\begin{equation}
h_{eff}(z)\left[
\begin{tabular}
[c]{l}%
$u_{\uparrow}$\\
$u_{\downarrow}$\\
$v_{\downarrow}$\\
$-v_{\uparrow}$%
\end{tabular}
\ \ \right]  =E_{n}\left[
\begin{tabular}
[c]{l}%
$u_{\uparrow}$\\
$u_{\downarrow}$\\
$v_{\downarrow}$\\
$-v_{\uparrow}$%
\end{tabular}
\ \ \ \ \right]
\end{equation}
with
\begin{equation}
h_{eff}(z)=\left[
\begin{array}
[c]{cc}%
H_{1D}(z) & \Delta\sigma_{0}\\
\Delta^{\ast}\sigma_{0} & -\sigma_{y}H_{1D}^{\ast}(z)\sigma_{y}%
\end{array}
\right]
\end{equation}
Using the above expression of $H_{1D}(z)$, one gets
\begin{equation}
h_{eff}(z)=h_{W}(z)+h_{C}(z)
\end{equation}
with%
\begin{align}
h_{W}(z)  &  =\left(  \frac{p_{z}^{2}}{2m}-\mu(z)+p_{z}\left(  \alpha
_{x}\sigma_{y}-\alpha_{y}\sigma_{x}\right)  \right)  \tau_{z}\nonumber\\
&  -\Delta\tau_{x}+E_{z}\sigma_{z} \label{MM}%
\end{align}
and%
\begin{equation}
h_{C}(z)=-\mu_{ac}\tau_{z} \label{HA}%
\end{equation}
In the following we disregard the term in $p_{z}^{2}/2m$ because we\ look for
solutions with a low $p_{z}$.

\subsection{A.3 Expressing $h_{W}(z)$ in a purely imaginary basis}

We define
\begin{align}
\alpha_{x}  &  =\alpha_{so}\cos(\theta_{so})\\
\alpha_{y}  &  =\alpha_{so}\sin(\theta_{so})
\end{align}
In the following, we work at first order in $p_{z}$ because we are only
interested in the low energy eigenstates of $h_{eff}(z)$. It is convenient to
express $h_{eff}(z)$ in a basis of self-adjoint operators. For this purpose we define%

\begin{equation}
R=\left[
\begin{array}
[c]{cccc}%
-\frac{i}{\sqrt{2}}e^{-\frac{i\theta_{so}}{2}} & \frac{1}{\sqrt{2}}%
e^{-\frac{i\theta_{so}}{2}} & 0 & 0\\
0 & 0 & \frac{1}{\sqrt{2}}e^{\frac{i\theta_{so}}{2}} & -\frac{i}{\sqrt{2}%
}e^{\frac{i\theta_{so}}{2}}\\
0 & 0 & \frac{1}{\sqrt{2}}e^{-\frac{i\theta_{so}}{2}} & \frac{i}{\sqrt{2}%
}e^{-\frac{i\theta_{so}}{2}}\\
-\frac{i}{\sqrt{2}}e^{\frac{i\theta_{so}}{2}} & -\frac{1}{\sqrt{2}}%
e^{\frac{i\theta_{so}}{2}} & 0 & 0
\end{array}
\right]  \label{R}%
\end{equation}
One can check%
\begin{align}
\widetilde{h}_{W}(z)  &  =R^{-1}h_{W}(z)R\nonumber\\
&  =\mu(z)\sigma_{y}\tau_{z}-i\hbar\alpha_{so}\tau_{x}\sigma_{z}\frac
{\partial}{\partial z}-E_{z}\sigma_{y}+\Delta\tau_{y}%
\end{align}%
\begin{equation}
\widetilde{h}_{C}(z)=R^{-1}h_{C}(z)R=-e\alpha_{c}V_{rms}\tau_{z}\sigma
_{y}(a+a^{\dag}) \label{hA}%
\end{equation}
Since $\widetilde{h}_{W}^{\ast}(z)=-\widetilde{h}_{W}(z)$ it is possible to
impose to all the zero energy eigenvectors%
\begin{equation}
\widetilde{\mathcal{\phi}}(z)=(u_{a}(z),u_{b}(z),u_{c}(z),u_{d}(z))^{t}%
\end{equation}
of $\widetilde{h}_{W}$ to be real. These eigenvectors correspond to operators%

\begin{align}
\gamma_{n}^{\dagger}  &  =%
{\textstyle\int}
dz^{\prime}(u_{a}(z^{\prime})\gamma_{a}(z^{\prime})+u_{b}(z^{\prime}%
)\gamma_{b}(z^{\prime})\nonumber\\
&  +u_{c}(z^{\prime})\gamma_{c}(z^{\prime})+u_{d}(z^{\prime})\gamma
_{d}(z^{\prime})) \label{op}%
\end{align}
with
\begin{align*}
\gamma_{a}(z)  &  =-\frac{i}{\sqrt{2}}e^{-\frac{i\theta_{so}}{2}}%
\psi_{\uparrow}^{\dag}(z)+\frac{i}{\sqrt{2}}e^{+\frac{i\theta_{so}}{2}}%
\psi_{\uparrow}(z)=\gamma_{a}^{\dag}(z)\\
\gamma_{b}(z)  &  =\frac{1}{\sqrt{2}}e^{-\frac{i\theta_{so}}{2}}\psi
_{\uparrow}^{\dag}(z)+\frac{1}{\sqrt{2}}e^{+\frac{i\theta_{so}}{2}}%
\psi_{\uparrow}(z)=\gamma_{b}^{\dag}(z)\\
\gamma_{c}(z)  &  =\frac{1}{\sqrt{2}}e^{\frac{i\theta_{so}}{2}}\psi
_{\downarrow}^{\dag}(z)+\frac{1}{\sqrt{2}}e^{-\frac{i\theta_{so}}{2}}%
\psi_{\downarrow}(z)=\gamma_{c}^{\dag}(z)\\
\gamma_{d}(z)  &  =-\frac{i}{\sqrt{2}}e^{\frac{i\theta_{so}}{2}}%
\psi_{\downarrow}^{\dag}(z)+\frac{i}{\sqrt{2}}e^{-\frac{i\theta_{so}}{2}}%
\psi_{\downarrow}(z)=\gamma_{d}^{\dag}(z)
\end{align*}
With this representation one can easily check that a zero energy normalized
eigenvector of $\widetilde{h}_{W}(z)$ corresponds to a Majorana bound state
(MBS) $\gamma_{n}^{\dagger}=\gamma_{n}$ with $\gamma_{n}^{2}=1/2$.

\subsection{A.4 Eigenstates of $\widetilde{h}_{W}(z)$}

\subsubsection{Uniform case}

In the case of a spatially constant $\mu$, assuming $\left\vert \mu\right\vert
<E_{z}$, the zero energy eigenstates of $\widetilde{h}_{W}(z)$ are $V_{k_{m}%
}^{+}\exp(k_{m}z)$, $V_{k_{m}}^{-}\exp(-k_{m}z)$, $V_{k_{p}}^{+}\exp(k_{p}z)$,
and $V_{k_{p}}^{-}\exp(-k_{p}z)$ with%

\begin{align}
k_{m}(\mu) &  =\frac{\Delta-\sqrt{E_{z}^{2}-\mu^{2}}}{\hbar\alpha_{so}}\\
k_{p}(\mu) &  =\frac{\Delta+\sqrt{E_{z}^{2}-\mu^{2}}}{\hbar\alpha_{so}}%
\end{align}%
\begin{align}
V_{m}^{+}(\mu) &  =(-\cos\phi(\mu),0,0,\sin\phi(\mu))^{t}\\
V_{m}^{-}(\mu) &  =(0,\cos\phi(\mu),\sin\phi(\mu),0)^{t}\\
V_{p}^{+}(\mu) &  =(\cos\phi(\mu),0,0,\sin\phi(\mu))^{t}\\
V_{p}^{-}(\mu) &  =(0,-\cos\phi(\mu),\sin\phi(\mu),0)^{t}%
\end{align}
and
\begin{equation}
\phi(\mu)=\arctan(\sqrt{\frac{E_{z}-\mu}{E_{z}+\mu}})
\end{equation}
Note that in order to find the above solutions, we have assumed that the term
in $p_{z}^{2}/2m$ is smaller than the other terms of the Hamiltonian
(\ref{MM}). This is valid provided
\begin{equation}
2m\alpha_{so}^{2}\gg\frac{\left(  \Delta-\sqrt{E_{z}^{2}-\mu^{2}}\right)
^{2}}{\min(\mu,E_{z},\Delta,\Delta-\sqrt{E_{z}^{2}-\mu^{2}})}%
\end{equation}
and%
\begin{equation}
2m\alpha_{so}^{2}\gg\frac{\left(  \Delta+\sqrt{E_{z}^{2}-\mu^{2}}\right)
^{2}}{\min(\mu,E_{z},\Delta)}%
\end{equation}
This criterion is largely satisfied in our work considering that the scale
$2m\alpha_{so}^{2}$ is typically huge ($\sim40~\mathrm{meV}$) in comparison
with $\Delta$ and $E_{z}$ ($\sim500~\mathrm{\mu eV}$).

\subsubsection{Non-uniform case, disregarding finite size effects}

In the main text, we study a nanowire with topological ($\mu=\mu_{1}<\mu_{c}$)
\ and non-topological ($\mu=\mu_{0}>\mu_{c}$) regions, with $\mu_{c}%
=\sqrt{E_{z}^{2}-\Delta^{2}}$ the chemical potential at which the bulk
topological transition occurs. We consider the $\mu(z)$ profile of the main
text, Figure 1.b. For $L_{T}\rightarrow+\infty$ and $L_{NT}\rightarrow+\infty
$, one has four MBSs appearing at $z=0$, $z=L_{T}$, $z=L_{T}+L_{NT}$,
$z=2L_{T}+L_{NT}$, with corresponding eigenfunctions $\widetilde
{\mathcal{\phi}}_{i}(z)$ such that $\widetilde{h}_{W}(z)\widetilde
{\mathcal{\phi}}_{i}(z)=0$, with $i\in\{1,2,3,4\}$. These four states
correspond to the Majorana operators $\gamma_{1}$, $\gamma_{2}$, $\gamma_{3}$
and $\gamma_{4}$ of the main text. One can check, for MBS 1:
\begin{align}
\widetilde{\mathcal{\phi}}_{1}(z  &  <0)=\frac{\mathcal{N}}{2}\Omega_{+}%
V_{m}^{+}(\mu_{0})\exp(k_{m}(\mu_{0})z)\nonumber\\
&  +\frac{\mathcal{N}}{2}\Omega_{-}V_{p}^{+}(\mu_{0})\exp(k_{p}(\mu_{0})z)
\end{align}%
\begin{equation}
\widetilde{\mathcal{\phi}}_{1}(z>0)=\mathcal{N}V_{m}^{+}(\mu_{1})\exp
(k_{m}(\mu_{1})z)
\end{equation}
and for MBS 2:
\begin{equation}
\widetilde{\mathcal{\phi}}_{2}(z<L_{T})=\mathcal{N}V_{m}^{-}(\mu_{1}%
)\exp(-k_{m}(\mu_{1})(z-L_{T}))
\end{equation}%
\begin{align}
\widetilde{\mathcal{\phi}}_{2}(z  &  >L_{T})=\frac{\mathcal{N}}{2}\Omega
_{+}V_{m}^{-}(\mu_{0})\exp(-k_{m}(\mu_{0})(z-L_{T}))\nonumber\\
&  +\frac{\mathcal{N}}{2}\Omega_{-}V_{p}^{-}(\mu_{0})\exp(-k_{p}(\mu
_{0})(z-L_{T}))
\end{align}
The vectors $V_{p}^{\pm}(\mu_{1})$ do not occur in these solutions because
their symmetry is not compatible with the solutions in the non-topological
phase (assuming we keep only normalizable
solutions)\cite{Lutchyn,Oreg,Sticlet}. Similarly, one has, for MBS 3:
\begin{equation}
\widetilde{\mathcal{\phi}}_{3}(z)=\widetilde{\mathcal{\phi}}_{1}%
(z-L_{T}-L_{NT})
\end{equation}
and for MBS 4:
\begin{equation}
\widetilde{\mathcal{\phi}}_{4}(z)=\widetilde{\mathcal{\phi}}_{2}%
(z-L_{T}-L_{NT})
\end{equation}
We have used above:%
\begin{equation}
\Omega_{\pm}=\frac{\sin\phi(\mu_{1})}{\sin\phi(\mu_{0})}\pm\frac{\cos\phi
(\mu_{1})}{\cos\phi(\mu_{0})}%
\end{equation}
and the normalization factor:%
\begin{equation}
\mathcal{N}=\sqrt{\frac{2\Delta\left(  \Delta^{2}+\mu_{0}^{2}-E_{z}%
^{2}\right)  \left(  \sqrt{E_{z}^{2}-\mu_{1}^{2}}-\Delta\right)  }{\hbar
\alpha_{so}(\mu_{0}-\mu_{1})\left(  \Delta\mu_{1}+\mu_{0}\sqrt{E_{z}^{2}%
-\mu_{1}^{2}}\right)  }}%
\end{equation}

\subsection{A. 5 Coupling between Majorana bound states for finite $L_{T}$ and
$L_{NT}$}

For finite values of $L_{T}$ and $L_{NT}$, we have to take into account a DC
coupling $\alpha_{ij}=%
{\textstyle\int}
\widetilde{\mathcal{\phi}}_{i}(z)\widetilde{h}_{W}(z)\widetilde{\mathcal{\phi
}}_{j}(z)$ between adjacent MBSs $i$ and $j$. We disregard the coupling
between non-adjacent bound states which is expected to be weaker. We use a
perturbation approach to calculate $\alpha_{ij}$, similar to
Ref.\cite{Shivamoggi}. We obtain the Hamiltonian $H_{wire}$ of the main text,
with $\epsilon$ and $\widetilde{\epsilon}$ real constants given by
$\alpha_{12}=\alpha_{34}=i\epsilon$ and $\alpha_{23}=i\widetilde{\epsilon}$.
One can check $\epsilon\simeq\lambda_{\epsilon}e^{k_{m}(\mu_{1})L_{T}}$ and
$\widetilde{\epsilon}\simeq\lambda_{\widetilde{\epsilon}}e^{-k_{m}(\mu
_{0})L_{NT}}$ with%

\begin{equation}
\lambda_{\epsilon}=2\zeta\sqrt{E_{z}^{2}-\mu_{1}^{2}}%
\end{equation}%
\begin{equation}
\lambda_{\widetilde{\epsilon}}=\zeta\left(  \sqrt{(E_{z}^{2}-\mu_{1}^{2}%
)}+\left(  (E_{z}^{2}-\mu_{0}\mu_{1})/\sqrt{E_{z}^{2}-\mu_{0}^{2}}\right)
\right)
\end{equation}
and%
\begin{equation}
\zeta=\Delta(\mu_{0}^{2}-\mu_{c}^{2})(\mu_{1}^{2}-\mu_{c}^{2})/(E_{z}(\mu
_{1}-\mu_{0})\vartheta)
\end{equation}
with%
\begin{equation}
\vartheta=E_{z}^{2}\mu_{0}+\mu_{1}\Delta^{2}-\mu_{0}\mu_{1}^{2}+\Delta(\mu
_{0}+\mu_{1})\sqrt{E_{z}^{2}-\mu_{1}^{2}}%
\end{equation}
The expression of $\widetilde{\epsilon}$ has been approximated using
\begin{equation}
\exp[-2L_{NT}\sqrt{E_{z}^{2}-\mu_{0}^{2}})/\hbar\alpha_{Z}]\ll1 \label{approx}%
\end{equation}

Cavity photons couple to MBSs due to $\widetilde{h}_{C}(z)$ defined in
Eq.(\ref{hA}). Again, it is sufficient to consider the coupling between
consecutive MBSs.\ The constants $\beta$ and $\widetilde{\beta}$ of the main
text correspond to $\beta(a+a^{\dag})=%
{\textstyle\int}
\widetilde{\mathcal{\phi}}_{1}(z)\widetilde{h}_{C}(z)\widetilde{\mathcal{\phi
}}_{2}(z)$ and $\widetilde{\beta}(a+a^{\dag})=%
{\textstyle\int}
\widetilde{\mathcal{\phi}}_{2}(z)\widetilde{h}_{C}(z)\widetilde{\mathcal{\phi
}}_{3}(z)$. Using (\ref{approx}), one finds the Hamiltonian $h_{int}$ of the
main text with%
\begin{equation}
\beta\simeq\lambda_{\beta}\frac{L_{T}}{l_{c}}\epsilon
\end{equation}%
\begin{equation}
\widetilde{\beta}\simeq\left(  \gamma_{\widetilde{\beta}}\frac{eV_{rms}}%
{\mu_{0}}+\lambda_{\widetilde{\beta}}\frac{L_{NT}}{l_{c}}\right)
\widetilde{\epsilon}%
\end{equation}%
\begin{equation}
\lambda_{\beta}=\frac{\mu_{1}}{\sqrt{E_{z}^{2}-\mu_{1}^{2}}}%
\end{equation}%
\begin{align}
\gamma_{\widetilde{\beta}} &  =\frac{E_{z}^{2}\mu_{0}(\mu_{1}-\mu_{0})}%
{(E_{z}^{2}-\mu_{0}^{2})(E_{z}^{2}-\mu_{0}\mu_{1}+\sqrt{(E_{z}^{2}-\mu_{0}%
^{2})(E_{z}^{2}-\mu_{1}^{2})})}\\
\lambda_{\tilde{\beta}} &  =\frac{\mu_{0}\left(  (E_{z}^{2}-\mu_{0}^{2}%
)\sqrt{E_{z}^{2}-\mu_{1}^{2}}+(E_{z}^{2}-\mu_{0}\mu_{1})\sqrt{E_{z}^{2}%
-\mu_{0}^{2}})\right)  }{(E_{z}^{2}-\mu_{0}^{2})(E_{z}^{2}-\mu_{0}\mu
_{1}+\sqrt{(E_{z}^{2}-\mu_{0}^{2})(E_{z}^{2}-\mu_{1}^{2})})}%
\end{align}
For the realistic parameters we consider, the dimensionless parameters
$\lambda_{\beta}$, $\gamma_{\widetilde{\beta}}$ and $\lambda_{\widetilde
{\beta}}$ are of the order of $1$ while $eV_{rms}/\mu_{0}\ll L_{NT}/l_{c}$.
This leads to
\begin{equation}
\widetilde{\beta}\simeq\lambda_{\widetilde{\beta}}\frac{L_{NT}}{l_{c}%
}\widetilde{\epsilon}%
\end{equation}

\section{Appendix B: Conductance of the Majorana nanowire}

The ensemble of the nanowire and the normal metal contact connected to MBS 1
can be described by a Hamiltonian $H_{wire}+H_{N}$ with\cite{Bolechetc}
\begin{equation}
H_{N}=%
{\textstyle\sum\limits_{p}}
\varepsilon_{p}c_{p}^{\dag}c_{p}+t(c_{p}^{\dag}-c_{p})\gamma_{1}%
\end{equation}
For simplicity, we assume that the coupling element $t$ between MBS 1 and the
contact is energy independent. Since the nanowire is tunnel coupled to a
grounded superconducting contact, a current can flow between this
superconducting contact and the normal metal contact, though the MBSs. The
conductance of the contact can be calculated as\cite{Flensberg}
\begin{equation}
G=(2e^{2}/h)%
{\textstyle\int}
d\varepsilon g_{0}(\varepsilon)\frac{df(\varepsilon-eV)}{d\varepsilon}%
\end{equation}
with $f(\varepsilon)=1+\exp(\varepsilon/k_{B}T)$ the Fermi function ,
$\Gamma=2\pi\nu_{0}\left\vert t\right\vert ^{2}$ the tunnel rate to between
the contact and MBS 1, $\nu_{0}$ the density of states in the contact and%
\begin{equation}
g_{0}=\frac{\Gamma^{2}\omega^{2}(\omega^{2}-4(\epsilon^{2}+\widetilde
{\epsilon}^{2}))^{2}}{\left\vert 16\epsilon^{4}+4\epsilon^{2}(i\Gamma
-2\omega)\omega+\omega(-i\Gamma+\omega)(\omega^{2}-4\epsilon^{2})\right\vert
^{2}}%
\end{equation}
Near the topological transition ($\varepsilon$ and $\widetilde{\varepsilon}$
finite), and if $\Gamma$ and $k_{B}T$ are small, the conductance $G$ displays
four peaks at $eV\simeq(\pm\hbar\omega_{e}\pm\hbar\omega_{o})/2$ which
correspond to the eigenenergies of Hamiltonian $H_{wire}$ of the main text. In
this case, the current flows between the superconducting contact and the
normal metal contact through the four MBSs which are coupled together
(Fig.3.a). Far from the topological transition ($\varepsilon\rightarrow0$), a
single zero energy resonance is visible, because MBS1, which is the only bound
state coupled directly to the normal metal contact, is disconnected from the
other MBSs. In this case, the current flows between the superconducting
contact and the normal metal contact through MBS 1 only (Fig.3.b).

\section{Appendix C: Kerr oscillator in the classical regime}

Following Ref.\cite{Yurke}, in the framework of the input/output
theory\cite{Walls}, the modulus $t_{cav}$ of the cavity transmission is given
by:%
\begin{equation}
t_{cav}=\frac{2\sqrt{\gamma_{in}\gamma_{out}}}{\sqrt{(\hbar(\omega
_{cav}-\omega_{RF})+2KE)^{2}+\gamma^{2}}}\label{tt}%
\end{equation}
with $\gamma_{in/out}$ the photonic transmission rate between the input/output
port and the cavity, $\gamma$ the total decoherence rate of cavity photons and
$E$ a semiclassical cavity photon number given by
\begin{equation}
E^{3}+\frac{\hbar\Delta\omega}{K}E^{2}+\frac{(\hbar^{2}\Delta\omega^{2}%
+\gamma^{2})}{4K^{2}}E=\frac{\gamma_{in}P_{1}^{in}}{K^{2}\hbar\omega_{RF}%
}\label{10}%
\end{equation}
with $\Delta\omega=\omega_{cav}-\omega_{RF}$. Above, $P_{1}^{in}$ and
$\omega_{RF}$ are the power and frequency of the input signal applied to the
cavity. From Eq.(\ref{10}), the cavity transmission becomes hysteretic for
$P_{1}^{in}>P_{1}^{crit}$ with%
\begin{equation}
P_{1}^{crit}=\frac{2}{3\sqrt{3}}\frac{\gamma^{3}}{\gamma_{in}\left\vert
K\right\vert }\hbar\omega_{cav}\label{crit}%
\end{equation}

\end{document}